\documentclass[conference]{IEEEtran}
\IEEEoverridecommandlockouts
\usepackage{cite}
\usepackage{amsmath,amssymb,amsfonts}
\usepackage{algorithm}
\usepackage{algpseudocode}
\usepackage{graphicx}
\usepackage{textcomp}
\usepackage{xcolor}
\usepackage{graphics}
\usepackage{svg}
\usepackage{epsfig}
\usepackage{balance}
\usepackage{mathtools}
\usepackage{hyperref}
\usepackage{makecell}
\def\BibTeX{{\rm B\kern-.05em{\sc i\kern-.025em b}\kern-.08em
    T\kern-.1667em\lower.7ex\hbox{E}\kern-.125emX}}

\DeclareMathOperator{\EX}{\mathbb{E}}

\begin{document}

\title{Best Memory Architecture Exploration 
under Parameters Variations 
accelerated with
Machine Learning 
}

\author{\IEEEauthorblockN{Antonios Tragoudaras\IEEEauthorrefmark{0}, Charalampos Antoniadis\IEEEauthorrefmark{0} and Yehia Massoud\IEEEauthorrefmark{0}
}
\IEEEauthorblockA{\IEEEauthorrefmark{0}Innovative Technologies Laboratories (ITL), \\
King Abdullah University of Science and Technology (KAUST),
Thuwal, Saudi Arabia \\ \{antonios.tragoudaras, charalampos.antoniadis, yehia.massoud\}@kaust.edu.sa
}
}

\maketitle
\begin{abstract}
The design of effective memory architecture is of utmost importance in modern computing systems. However, the design of memory subsystems is even more difficult today because process variation and modern design techniques like dynamic voltage scaling make performance metrics for memory assessment be treated as random variables instead of scalars at design time. Most of the previous works have studied the design of memory design from the yield analysis perspective leaving the question of the best memory organization on average open. Because examining all possible combinations of design parameter values of a memory chip would require prohibitively much time, in this work, we propose Best Arm Identification (BAI) algorithms to accelerate the exploration for the best memory architecture on average under parameter variations. Our experimental results demonstrate that we can arrive at the best memory organization 99\% of the time in $\times$5 faster than an exhaustive search of all possible conditions.
\end{abstract}


\section{Introduction}


Memory subsystems play a substantial role in the overall performance of modern smart computing systems that execute those deep-learning models. Therefore the exploration of different memory architectures is essential in the overall system design. Because of the aggressive technology scaling and new architectural design techniques such as Dynamic Voltage Scaling (DVS), transistor design parameters and the supply voltage cannot be assumed constants but random variables. Because of that, memory performance metrics, e.g., access time and power, are also random variables at design time. So, a memory architecture may be the best for specific values of the design parameters but worse than the other memory architectures for different values of the design parameters. Therefore, the problem is to discover the best memory architecture on average. 

Previous works in memory design under parameter variations have mostly paid attention to modeling the yield (percentage of memory chips that meet the performance constraints) by taking into account all architectural correlations between the different memory components~\cite{IMEC} and yield enhancement~\cite{myDATE15}. The previous works discover the Probability Density Function (PDF) of the performance metrics using Monte Carlo (MC) simulations. MC is more accurate than distribution propagation methods, and the preferable method in Electronic Design Automation (EDA) for statistical analysis~\cite{myDATE13}\cite{myDATE18} in general, because it avoids simplistic assumptions about the distributions made for efficient distribution propagation through complex systems, like memory circuits. 
Moreover, these works employ variance-reduction statistical techniques to reduce the number of MC simulations but again adjusted for yield estimation and not for finding the distribution mean of the performance metric of interest. More specifically, they sample circuit design parameter values that contribute more to yield estimation, i.e., the tails of the distributions of the parameters ignoring the bulk of the distribution where the distribution's mean value is located. Apparently, the estimation of the distribution's mean value of the metric of interest in the memory circuit suffers from the same drawbacks mentioned before for yield estimation. Thus, exhaustive MC is the only viable method for finding the best architecture on average under parameter variations. However, an exhaustive MC requires prohibitively much time.

Best Arm Identification (BAI) is the problem of finding the most rewarding arm to pull under a limited budget. All available arms/bandits are initially pulled to obtain a reward according to some random unknown distribution probability. A score function is then calculated for each arm to determine which will be selected next (the one with the highest score). This procedure is repeated until the available budget is exhausted, followed by the final selection of the most rewarding arm. The BAI problem has found application in different areas, e.g., the design of clinical trials~\cite{Villar2015}, recommendation systems of news and goods~\cite{Li2010}, the binding pose prediction between ligands and proteins~\cite{protein_ligand}, and the game of Go~\cite{Coulom2006},~\cite{Silver2016}. In this paper, we employ BAI algorithms, for the first time to the best of our knowledge, to reduce the number of required MC simulations in exploring the best memory architecture on average under process and voltage variations. More specifically, we represent a candidate memory architecture with an arm, the act of pulling an arm with a simulation (MC trial) to obtain performance metrics of the candidate memory architecture (for random memory chip design parameter values), and a reward with the cost of the candidate architecture based on the previous performance metrics.


The rest of the paper is organized as follows: Section~\ref{sec:Background} presents the problem formulation and explains a memory system structure, the cost function we consider for memory architecture evaluation, and the power and delay models we use in this work. Section~\ref{sec:Proposed BAI Algorithms} explains the proposed BAI algorithms to accelerate the exploration for the best memory architecture on average under process and voltage variations. Finally, section~\ref{sec:Experimental Results} demonstrates the reduced simulations required when using the proposed BAI algorithms compared to exhaustive MC simulations, while conclusions are drawn in Section~\ref{sec:Conclusion}.

\section{Background}
\label{sec:Background}

\subsection{Memory Organization}
\label{subsec:mem_organization}

\begin{figure}[h!]
  \centering
  \includegraphics[width=\linewidth]{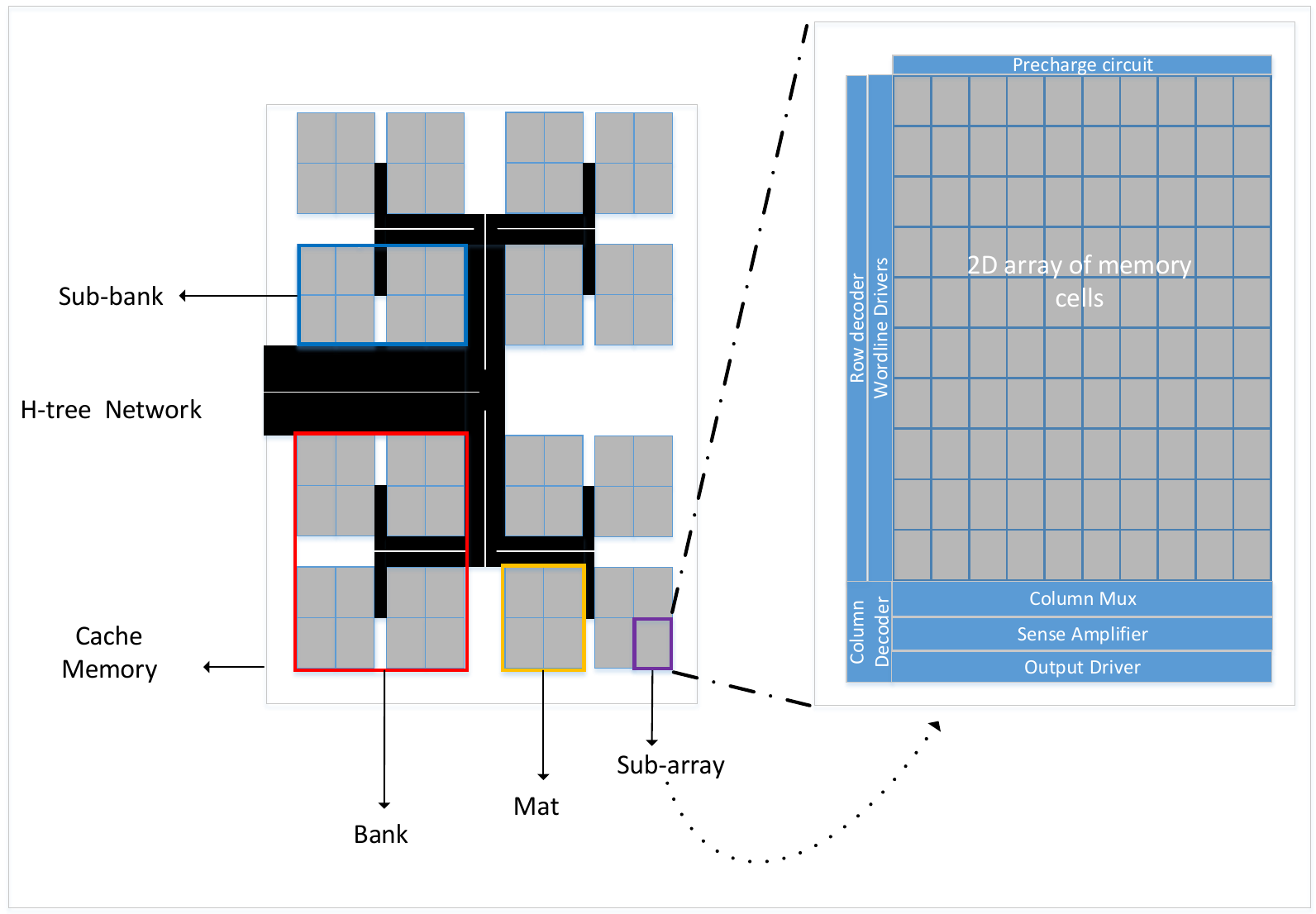}
  \caption{Cache Organization. The figure illustrates a cache memory consisting of 4 banks where each bank is divided into 2 sub-banks and each sub-bank is divided into 2 mats.}
  \label{figure:fig1}
\end{figure}

As shown in Figure~\ref{figure:fig1}, memories and caches are storage arrays with a hierarchical organization. In the remainder of the paper, we study such a memory structure that is also assumed in CACTI~\cite{CACTI}; however, all applied techniques are applicable to other memory structures as well. A cache memory consists of $N_{\mathrm{banks}}$ identical banks at the highest level of abstraction. Multiple banks may be accessed simultaneously because each bank has its own address/data bus. A bank consists of $N_{\mathrm{subbanks}}$ of identical sub-banks that are activated sequentially with each access. In turn, each sub-bank contains a number of identical mats. A mat is a self-contained, compact memory array composed of four identical sub-arrays, with $N_{\mathrm{rows}}$ rows and $N_{\mathrm{cols}}$ columns, and accompanying predecoding logic, with each sub-array being a two-dimensional matrix of memory cells and associated peripheral circuitry. Each mat holds a portion of a word in one of its four sub-arrays; during cache access, all mats in a sub-bank are activated to form the whole word. H-tree distribution networks are often used to deliver address and data to mats.



\subsection{Cost Function for Memory Architecture Evaluation}
\label{subsec:Cost function}
Based on CACTI cost function in order to evaluate different memory architectures for different performance metrics we consider the following cost function: 
\begin{equation}
Cost(Arch) = w_{t}*\widetilde{T}_{acc}(Arch) + w_{P_{dyn}}*\widetilde{P}_{dyn}(Arch),
\end{equation}
\noindent
where $0 \leq w_{t}, w_{dyn} \leq 1$ and $w_{t} + w_{dyn} = 1$, i.e., user-defined weights adjusting the importance of access Time ($T_{acc}$) and dynamic Power ($P_{dyn}$) for the candidate architecture $Arch$ = $\lbrace N_{banks}, N_{subbanks}, N_{rows}, N_{cols} \rbrace$ in $Cost$, $\widetilde{T}_{acc}(Arch) = \frac{T_{acc}(Arch)}{minT_{acc}}$ and $\widetilde{P}_{dyn}(Arch) = \frac{P_{dyn}(Arch)}{minP_{dyn}}$ are the normalized access Time and dynamic Power with $minT_{acc}$ and $minP_{dyn}$ being the minimum access Time and dynamic Power, respectively, anytime during the search for the best architecture $Arch^{*}$.

 \subsection{Delay and Dynamic Power Modeling}
\label{subsec:Delay and dynamic Power modeling}

For modeling access time and dynamic power modeling under process and voltage variations, we use the tool proposed in~\cite{myDATE15} as an extension of CACTI. Access time results from the addition of gate delays in the memory components. For the gate delay modeling, we employ the RC model. RC delay approximations entail the estimation of the equivalent resistance, a.k.a the effective resistance $R_{eff}$. The effective resistance is the ratio of the drain-source voltage $V_{DS}$ to the drain-source current $I_{DS}$ of the driving transistor averaged across the switching interval of interest~\cite{Weste} as shown below:

\begin{equation}\label{eq:Reff}
R_{eff}\ =\ \frac{V_{DD}}{I_{eff}},
\end{equation}

\noindent
with 

\begin{equation}
I_{eff}\ =\ \frac{I_{H}+I_{L}}{2},\label{eq:Ieff},
\end{equation}

\noindent
where $I_{H}=I_{DS}(V_{GS}=V_{DD}, V_{DS}=\frac{V_{DD}}{2})$ and $I_{L}=I_{DS}(V_{GS}=\frac{V_{DD}}{2},V_{DS}=V_{DD})$.

For $I_{DS}$ modeling we use the alpha-power law model~\cite{Sakurai} shown below 

\begin{equation}\label{eq:alpha.Current.Model}
\begin{split}
I_{DS} =
\begin{cases}
0,\ &if\ V_{GS} \leq V_{th}  \\
\frac{W}{L_{eff}} \frac{P_{c}}{P_{u}}(V_{GS}-V_{th})^{\alpha/2}V_{DS},\ &if\ V_{DS} < V_{d0}\\ 
\frac{W}{L_{eff}}P_{c}(V_{GS}-V_{th})^\alpha,\ &if\ V_{DS} \geq V_{d0},
\end{cases}
\end{split}
\end{equation}

\noindent
which is more suitable for lower technology nodes because it captures short channel effects better compared to the Shockley model. In this equation $V_{GS}$ and $V_{DS}$ are the gate-source and the drain-source voltages, respectively, $P_{c}$ and $P_{u}$ are constants and $V_{d0}$ is given by:

\begin{center}
$V_{d0}\ =\ P_{u}(V_{GS}-V_{th})^{\alpha/2}$.
\end{center}

\noindent
Moreover, we determined $\alpha = 1.3$ to be in perfect agreement with ITRS~\cite{ITRS} reports after several experiments for the 65nm technology node.  

In addition, the dynamic power of memory is the sum of the dynamic
power of the logic gates on the path 
activated during memory access. For modeling the dynamic
power of logic gates, we use the following model:

\begin{equation}
P_{dyn}\ =\ 0.5*C_{total}*V_{DD}^{2},
\end{equation}

\noindent
where $C_{total}$ is the total load seen by the gate output. 

\subsection{Problem Formulation}
\label{subsec:Problem Formulation}

So, we consider the following optimization problem in this work:

\begin{equation}
\begin{aligned}
\underset{Arch}{argmin} 
\quad &\underset{\boldsymbol\theta \sim\Theta}{\EX}[Cost(Arch_{\boldsymbol\theta})], \\
\textrm{s.t.} \quad &\underset{\boldsymbol\theta \sim\Theta}{\EX}[T_{acc}(Arch_{\boldsymbol\theta})] < T_{acc}^{target} \quad \text{and} \\ 
&\underset{\boldsymbol\theta \sim\Theta}{\EX}[P_{dyn}(Arch_{\boldsymbol\theta})] < P_{dyn}^{target},
\end{aligned}
\end{equation}

\noindent
where $Arch$ is the unknown configuration of the memory architecture, $T_{acc}^{target}$ is the access time target, $P_{dyn}^{target}$ is the dynamic power target, and $\boldsymbol\theta$ drawn from $\Theta$ distribution represents the random variables corresponding to the parameters of the different transistors. Essentially, we are looking for the memory architecture that minimizes the average cost under random values for the transistors' parameters in the memory circuit. Because Monte Carlo simulations can more accurately catch the problem's stochasticity than probability distribution propagation methods but at the same time require prohibitively much more time, as we explained in the Introduction, we introduce three BAI algorithms in section \ref{sec:Proposed BAI Algorithms} to reduce these Monte Carlo simulations but without sacrificing the accuracy of the results.

\section{Proposed BAI Algorithms}
\label{sec:Proposed BAI Algorithms}

\begin{figure}
    \centering
    \includegraphics[width=0.4\textwidth]{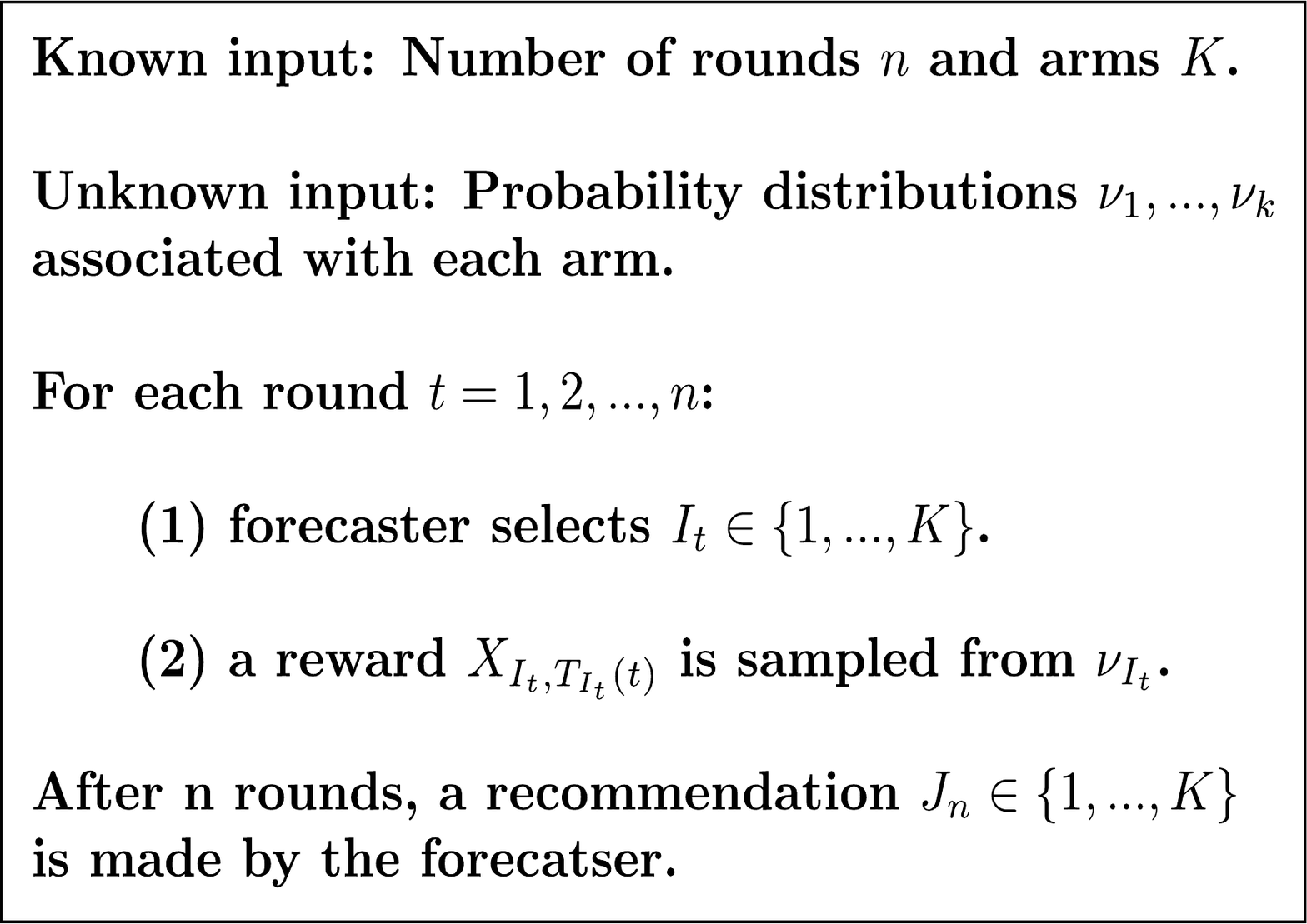}
    \caption{Overview of the multi-armed bandit problem} 
    \label{fig:problem_formulation}
\end{figure}

\noindent In this section, we delineate the intuition behind each BAI algorithm used in our study. Let us begin by defining the general BAI formulation (see Fig.~\ref{fig:problem_formulation}). 
Given a set $\mathit{A} = \mathit{\{1,..., K\}}$ of a finite number of arms and $\textit{n}$ number of rounds,
there are unknown probability distributions $\nu_{i}$, $i = 1 \ldots K$ associated with the reward obtained after pulling arm $i$.
At each round $\mathit{t} = \mathit{\{1,..., n\}}$, the forecaster makes an arm selection $\mathit{I_t}$ from set $\mathit{A}$, and observes a reward $\mathit{\nu_{I_t}}$, which is 
uncorrelated with previous observations. For every arm $\mathit{i}$ and after each round $\mathit{t}$, the forecaster keeps track of the number of times $\mathit{T_{i}(t)}$ arm $\mathit{i}$ has been pulled and the sequence of associated rewards ${\mathit{X_{i,1}}, ..., \mathit{X_{i,T_{i}(t)}}}$ at each round. $\mathit{\hat{X}_{i,s}}$ $= \sum_{t=1}^{s} X_{i,t}$ denotes the mean of the sampled rewards after $\mathit{s}$ pulls of arm $\mathit{i}$. After the last round $\mathit{n}$, the forecaster makes an arm recommendation $\mathit{J_n}\in\mathit{A}$ which corresponds to the arm with the maximum empirical mean $\mathit{\hat{X}_{i,T_{i}(n)}}$.\par
The BAI formulation described in the previous paragraph is the fixed budget setting approach ($n$ rounds). At this setting to the best of our knowledge, the UCB-E \cite{audibert2010}, and UGapE\cite{Gabillon2012,gabillon2011multi} achieve the lowest bounds of the probability of error, namely the probability the forecaster has selected an arm $\mathit{J_n}$ which is not the optimal after $\mathit{n}$ rounds. The limitation of both algorithms lies in the upper bound probability of error 
requiring a precise estimate of the \textit{complexity} term $\mathit{H}$ 
found in all BAI algorithms.
That is because this calculation depends primarily on selecting a user-defined exploration parameter and secondarily on the rewards distribution of the arm, both unknown to the forecaster. The overestimation of complexity $\mathit{H}$ imposes a low exploration policy that is proven to lead to incorrect arm recommendation in some cases\footnote{For instance, the actual best arm has given an unusually low reward.}. On the other hand, complexity underestimation results in over-exploring sub-optimal arms, obstructing the search from developing a decent estimation for the mean rewards of the best-performing arms. In order to avoid these pitfalls, we decided to utilize an algorithm free of user-defined exploration parameters, called Successive Rejects\cite{audibert2010} requiring no previous knowledge of the complexity term. Similarly, we employ the adaptive variants of UCB-E and UGapE to auto-tune their exploration parameters appropriately to derive a decent estimation for their complexity terms $\mathit{H}$. 
\subsection{Successive Rejects - Exploration Parameter Free Algorithm}

\begin{algorithm}
 \scriptsize
\caption{\scriptsize SR algorithm. Given $A_1 = \{1,...,K\}$, $\overline{\log}(K) = \frac{1}{2} +  \sum_{i=2}^{K} \frac{1}{i}$,\newline $n_{0} = 0$ and for $k \in \{1,...,K-1\}$: $n_{k} = \lceil \frac{1}{\overline{\log}(K)}\frac{n-K}{K+1-k} \rceil$}
\label{alg:SR_algo_description}
\begin{algorithmic}[1]
\For{phase $k$ = 1,...,K-1}
	\For{arm $i\in A_{k}$}
        \State{Draw arm $i$ for $n_{k}-n_{k-1}$ rounds \newline
        \hspace*{3em} \& sample the rewards.}
	\EndFor
	\State{Update $A_{k+1}=A_{k}\backslash argmin_{i\in A_{k}}\hat{X}_{i,n_{k}} $ \newline
	\hspace*{1em} (Remove the worst Arm in $A_{k}$)  }
\EndFor
\State \Return{$J_{N}\in A_{k}$ (the surviving Arm in $A_{K}$)}
\end{algorithmic}
\end{algorithm}

The Successive Reject algorithm imposes a strategy that splits the time into $K-1$ phases instead of rounds. At the end of each phase, the seemingly worst arm is eliminated. Then, during the evaluation of a phase, all the remaining arms are pulled with a "fair" frequency of times, controlled by $n_{k}$ (see Algorithm~\ref{alg:SR_algo_description}). After completing all phases, the only remaining component in $A_{K}$ is the forecaster's recommendation as the best arm.

\subsection{Adaptive Upper Confidence Bound Exploration}

\begin{algorithm}
\scriptsize
\caption{\scriptsize Auto UCB-E algorithm. Given $\overline{\log}(K) = \frac{1}{2} + \sum_{i=2}^{K} \frac{1}{i}$, $n_{0} = 0$ and for $k \in \{1,...,K-1\}$: $n_{k} = \lceil \frac{1}{\overline{\log}(K)}\frac{n-K}{K+1-k} \rceil$,\newline$t_{0} = 0$, $t_1 = Kn_{1}$ and for $k > 1$, $t_{k}=n_1+...+n_{k-1} + (K-k+1)n_k$,\newline$\hat{\Delta}_{i}(t)=(\underset{1\leq j\leq K}{max} \hat{X}_{j,T_{j}(t_k)})-\hat{X}_{i,T_{i}(t_K)}$ and $<i>$ in an ordering that satisfies $\hat{\Delta}_{<1>} \leq ...\leq \hat{\Delta}_{<k>}$. For arm $i \in \{1,..,K\}$ and $b'>0$, let $S_{i,{s}}(b') = \hat{X}_{i,s} + \sqrt{\frac{b'}{s}}$, where $s \geq 1$ and $S_{i,s}(0) = +\infty$. $n$ total number of rounds.}
\label{alg:UCBe_auto}
\begin{algorithmic}[1]
\For{phase $k$ = 0,...,K-1}
    \If{k=0}
        \State{$\hat{H}_{b,k} = K$}
    \Else (Complexity Calculation at phase $K$)
        \State{$\hat{H}_{c,k} = \underset{K-k+1\leq i\leq K}{max}i\hat{\Delta}_{<i>}^{-2}$}
    \EndIf

    \For{round $t=t_{k}+1,...,t_{k+1}$}(Pull arms for those rounds)
        \State{Select Arm $I_{t}\in \underset{i\in \{1,...,K\}}{argmax}S_{i,T_{i}(t-1)}(\frac{n}{\hat{H}_{b,k}})$}
        \State{Get reward $X_{I_{t},T_{I_t}(t-1)+1}$}
    \EndFor
\EndFor
\State \Return{$J_{n}\in \underset{i\in \{1,..,K\}}{argmax}\hat{X}_{i,T_{i}(n)}$}
\end{algorithmic}
\end{algorithm}


\noindent The original UCB-E strategy was designed as a highly-exploring policy. However, the efficacy of its exploration strategy is heavily dependent on a user-defined exploration parameter, which can lead to over/underestimation of the complexity term if it is not appropriately fine-tuned beforehand. The adaptive UCB-E is a variant of the "offline" UCB-E method, alleviating the need to carefully select the value of exploration parameter $b$ by calculating proper estimates for the problem complexity term $H_b$. Algorithm~\ref{alg:UCBe_auto} describes the implementation of the adaptive UCB-E algorithm, as was initially proposed in \cite{audibert2010}, capable of demonstrating a sufficient high exploring policy for the discovery of optimal arms. The adaptive UCB-E variant has adopted the strategy of using phases as in SR, which estimates the complexity $\hat{H}_{b}$ at each given phase achieving fine-tuning for exploration parameter $b$.  

\subsection{Adaptive Uniform Gap-based Exploration}

\begin{algorithm}
\scriptsize
\caption{\scriptsize Uniform GapE algorithm.\newline a: exploration parameter; $a > 0$, n:total number of rounds,\newline$\beta_{k}(t-1)$: confidence interval, $\beta_{k}(t-1) = \sqrt{\frac{a}{T_{k}(t-1)}}$,\newline$U_{k}(t), L_{k}(t)$ are the upper and lower bounds of high probability where for arm $k\in A$ and round $t$: $U_{k}(t) = \hat{X}_{k,T_{k}(t-1)} + \beta_{k}(t-1)$, $L_{k}(t) = \hat{X}_{k,T_{k}(t-1)} - \beta_{k}(t-1)$}
\label{alg:UGapE_conv}
\begin{algorithmic}[1]
\For{$i$ = 1,...,K}
    \State{Pull arm $i$ and observe its reward $X_{i,1}$} 
\EndFor
\For{$t=K+1,...,n$}
    \For{$k=1,...,K$} (Calculate index $B_k$ of arm $k$)
        \State{$B_k(t) = \underset{i\neq k}{max}U_{i}(t) - L_{k}(t)$}
    \EndFor
    \State{$l_{t} =  arg\underset{k\in A}{min}B_{k}(t)$}
    \State{$u_{t} = arg\underset{k\in A\backslash l_t}{max}U_{j}(t)$}
    \State{Pick Arm $I_{t}\in arg\underset{k\in \{l_{t},u_{t}\}}{max}\beta_{k}(t-1)$}
    \State{Get reward $X_{I_{t},T_{I_{t}}(t-1)+1}$}
\EndFor
\State \Return{$J_{n}\in argmax_{i\in A}\hat{X}_{i, T_{i}(n)}$}
\end{algorithmic}
\end{algorithm}

The details of the original Uniform Gap-based exploration approach are illustrated in Alg.\ref{alg:UGapE_conv}. The key feature of the algorithm lies in the selection step, where UGapE has two arm options. The first one is to pull the arm $l_t$ with the minimum index $B_k$, then remove $l_t$ from the given snapshot of available arms in $A$. Alternatively, the arm $u_{t}$ with the highest high-probability upper bound $U_{k}t$ from the new snapshot of arms $A$, after discarding $l_t$, is chosen. The one with the highest uncertainty term $\beta(t-1)$ is selected between these two arms. The user-defined exploration parameter $a$ affects this selection considerably. The adaptive version of UGapE defines a suitable complexity estimate $\hat{H}_{a}(t)$ so that the exploration parameter can be derived as $a = \frac{n-k}{\hat{H}_{a}(t)}$ for round $t$, avoiding the drawbacks that come with the manual exploration parameter selection approach. In our implementation, we adopt the same approximation for complexity $H_{a}(t)$ as the one proposed by the authors of \cite{gabillon2011multi}.
\section{Experimental Results}
\label{sec:Experimental Results}

In this section, we evaluate the effectiveness of BAI algorithms in discovering optimal memory architectures on average under variations in memory circuit design parameters. We examined $1$MB memory architectures like those described in Subsection\ref{subsec:mem_organization},
considering the following parameter variations:

\begin{itemize}
    \item We assume different voltages for the peripheral circuitry and the memory cells. More specifically, we examine 4 possible voltage levels, i.e., 0.8V, 0.9V, 1.0V, and 1.1V, for the memory peripheral circuitry and cells resulting in 16 combinations.
    \item Transistor threshold voltages are drawn from a normal distribution with mean value the nominal value $V_{th_{0}} = 0.195V$ in 65nm~\cite{ITRS}, and a standard deviation $\sigma_{V_{th_{0}}} = 0.15 \times V_{th_{0}}$.
\end{itemize}

For our experiments, we assume $N$ = 17166 candidate memory architectures of 1 MB. Moreover, we consider uniform sampling (or exhaustive Monte Carlo) as our baseline method to identify the ``golden" memory architecture in our test cases, i.e., all candidate memory architectures are simulated with different sets of random supply and threshold voltage values for the peripheral circuitry/cells and the transistors, respectively in every round until the available budget is exhausted. We verified that a budget of $n$ = 100 is sufficient to capture the effects of supply and threshold voltage variations in the performance of candidate memory architectures. To compare the efficacy of the proposed BAI algorithms, we count the total number of simulations
required so that the BAI algorithms recommend the same architecture with the uniform sampling method (we will refer to that architecture as the ``correct" architecture). More specifically, we compute the probability of ``correct" memory architecture recommendation by measuring the ratio of ``correct" memory architecture recommendations for 1000 runs of each BAI algorithm. Tables~\ref{access_time_corr_prob}-\ref{fifty_fifty_corr_prob} assess the effectiveness of each BAI algorithm in recommending the ``correct" memory architecture for different user-defined performance metrics by adjusting the weights of the cost function (see Subsection~\ref{subsec:Cost function}). It is worth noticing that the average probability (last column of Tables~\ref{access_time_corr_prob}-\ref{fifty_fifty_corr_prob}) for all BAI algorithms recommending the ``correct" best memory architecture in all test cases is above 90\% after 20 $\times$ $N$ simulations, i.e., only one-fifth of the total number of simulations uniform sampling (or an exhaustive MC search) would require (recall that the total number of simulation with uniform sampling is 100 times $N$). The adaptive variants of UCB-E and UGapE are the winning strategies, having achieved very similar average probabilities of ``correct" recommendations above 99\% in all test cases for 20 $\times$ $N$ simulations. Our experiments indicate that integrating BAI strategies into the exploration for the best memory architecture on average under memory-chip design parameter variations would lead to significantly fewer simulations than an exhaustive search.

\begin{table}[!h]
    \centering
    \caption{Correct memory recommendation on the test case with $100\%$ focus on access time.}
    \label{access_time_corr_prob}
    \tabcolsep=0.15cm
    \begin{tabular}{|cccccc|}
        \hline 
        \thead{Simulations}&\thead{Uniform\\(Baseline)}&\thead{SR}&\thead{UCB-E\\auto}&\thead{UGapE\\auto}&\thead{BAI\\Average}\\
        \hline
         $10\times N$&24.1\%&36.3\%&57.2\%&36.7\%&43.4\%\\ 
         $15\times N$&27.8\%&63.7\%&72.9\%&51.9\%&62.8\%\\
         $20\times N$&27.6\%&78.6\%&97.6\%&100\%&92\%\\
        \hline
    \end{tabular}
\end{table}

\begin{table}[!h]
    \centering
    \caption{Correct memory organization recommendation on the test case with $100\%$ focus on dynamic power.}
    \label{dyn_power_corr_prob}
    \tabcolsep=0.15cm
    \begin{tabular}{|cccccc|}
        \hline 
        \thead{Simulations}&\thead{Uniform\\(Baseline)}&\thead{SR}&\thead{UCB-E\\auto}&\thead{UGapE\\auto}&\thead{BAI\\Average}\\
        \hline
         $10\times N$&31.7\%&47.8\%&76.2\%&55.7\%&59.9\%\\ 
         $15\times N$&39.8\%&94.4\%&97.8\%&100\%&97.4\%\\
         $20\times N$&40.6\%&99.2\%&100\%&100\%&99.73\%\\
        \hline
    \end{tabular}
\end{table}

\begin{table}[!h]
    \centering
    \caption{Correct memory organization recommendation on the test case with equal focus on power \& time.}
    \label{fifty_fifty_corr_prob}
    \tabcolsep=0.15cm
    \begin{tabular}{|cccccc|}
        \hline 
        \thead{Simulations}&\thead{Uniform\\(Baseline)}&\thead{SR}&\thead{UCB-E\\auto}&\thead{UGapE\\auto}&\thead{BAI\\Average}\\
        \hline
         $10\times N$&34.8\%&55.6\%&74.7\%&54.8\%&61.7\%\\ 
         $15\times N$&38.7\%&98.6\%&99.7\%&94\%&97.4\%\\
         $20\times N$&40.5\%&100\%&100\%&99.8\%&99.93\%\\
        \hline
    \end{tabular}
\end{table}

\section{Conclusion}
\label{sec:Conclusion}

In this study, we have developed a memory circuit design parameters variation-aware framework for making accurate recommendations on discovering optimal memory architectures on average utilizing Best Arm Identification algorithms. Our findings demonstrated that our approach could achieve an average recommendation accuracy of 99\% across three test cases while minimizing the search cost by one-fifth compared to exhaustive MC simulations.

\balance

\bibliographystyle{IEEEtran}
\bibliography{bibliography}

\begin{thebibliography}{10}
\providecommand{\url}[1]{#1}
\csname url@samestyle\endcsname
\providecommand{\newblock}{\relax}
\providecommand{\bibinfo}[2]{#2}
\providecommand{\BIBentrySTDinterwordspacing}{\spaceskip=0pt\relax}
\providecommand{\BIBentryALTinterwordstretchfactor}{4}
\providecommand{\BIBentryALTinterwordspacing}{\spaceskip=\fontdimen2\font plus
\BIBentryALTinterwordstretchfactor\fontdimen3\font minus
  \fontdimen4\font\relax}
\providecommand{\BIBforeignlanguage}[2]{{%
\expandafter\ifx\csname l@#1\endcsname\relax
\typeout{** WARNING: IEEEtran.bst: No hyphenation pattern has been}%
\typeout{** loaded for the language `#1'. Using the pattern for}%
\typeout{** the default language instead.}%
\else
\language=\csname l@#1\endcsname
\fi
#2}}
\providecommand{\BIBdecl}{\relax}
\BIBdecl

\bibitem{IMEC}
P.~Zuber \emph{et~al.}, ``Statistical sram analysis for yield enhancement.'' in
  \emph{DATE}, 2010, pp. 57--62.

\bibitem{myDATE15}
C.~Antoniadis, G.~Karakonstantis, N.~Evmorfopoulos, A.~Burg, and G.~Stamoulis,
  ``On the statistical memory architecture exploration and optimization,'' in
  \emph{2015 Design, Automation \& Test in Europe Conference \& Exhibition
  (DATE)}, 2015, pp. 543--548.

\bibitem{myDATE13}
A.~Cevrero, N.~Evmorfopoulos, C.~Antoniadis, P.~Ienne, Y.~Leblebici, A.~Burg,
  and G.~Stamoulis, ``Fast and accurate ber estimation methodology for i/o
  links based on extreme value theory,'' in \emph{2013 Design, Automation \&
  Test in Europe Conference \& Exhibition (DATE)}, 2013, pp. 503--508.

\bibitem{myDATE18}
C.~Antoniadis, D.~Garyfallou, N.~Evmorfopoulos, and G.~Stamoulis, ``Evt-based
  worst case delay estimation under process variation,'' in \emph{2018 Design,
  Automation \& Test in Europe Conference \& Exhibition (DATE)}, 2018, pp.
  1333--1338.

\bibitem{Villar2015}
S.~S. Villar, J.~Bowden, and J.~M.~S. Wason, ``Multi-armed bandit models for
  the optimal design of clinical trials: Benefits and challenges.''
  \emph{Statistical science : a review journal of the Institute of Mathematical
  Statistics}, vol. 30 2, pp. 199--215, 2015.

\bibitem{Li2010}
\BIBentryALTinterwordspacing
L.~Li, W.~Chu, J.~Langford, and R.~E. Schapire, ``A contextual-bandit approach
  to personalized news article recommendation,'' \emph{CoRR}, vol.
  abs/1003.0146, 2010. [Online]. Available:
  \url{http://arxiv.org/abs/1003.0146}
\BIBentrySTDinterwordspacing

\bibitem{protein_ligand}
\BIBentryALTinterwordspacing
K.~Terayama, H.~Iwata, M.~Araki, Y.~Okuno, and K.~Tsuda, ``{Machine learning
  accelerates MD-based binding pose prediction between ligands and proteins},''
  \emph{Bioinformatics}, vol.~34, no.~5, pp. 770--778, 10 2017. [Online].
  Available: \url{https://doi.org/10.1093/bioinformatics/btx638}
\BIBentrySTDinterwordspacing

\bibitem{Coulom2006}
R.~Coulom, ``Efficient selectivity and backup operators in monte-carlo tree
  search,'' in \emph{Proceedings of the 5th International Conference on
  Computers and Games}, ser. CG'06.\hskip 1em plus 0.5em minus 0.4em\relax
  Berlin, Heidelberg: Springer-Verlag, 2006, p. 72–83.

\bibitem{Silver2016}
D.~Silver, A.~Huang, C.~J. Maddison, A.~Guez, L.~Sifre, G.~van~den Driessche,
  J.~Schrittwieser, I.~Antonoglou, V.~Panneershelvam, M.~Lanctot, S.~Dieleman,
  D.~Grewe, J.~Nham, N.~Kalchbrenner, I.~Sutskever, T.~Lillicrap, M.~Leach,
  K.~Kavukcuoglu, T.~Graepel, and D.~Hassabis, ``Mastering the game of {Go}
  with deep neural networks and tree search,'' \emph{Nature}, vol. 529, no.
  7587, pp. 484--489, Jan. 2016.

\bibitem{CACTI}
S.~J.~E. Wilton \emph{et~al.}, ``Cacti: An enhanced cache access and cycle time
  model,'' \emph{IEEE JSSC}, vol.~31, pp. 677--688, 1996.

\bibitem{Weste}
N.~Weste \emph{et~al.}, \emph{CMOS VLSI Design: A Circuits and Systems
  Perspective}, 4th~ed.\hskip 1em plus 0.5em minus 0.4em\relax Addison-Wesley
  Publishing Company.

\bibitem{Sakurai}
T.~Sakurai \emph{et~al.}, ``Alpha-power law mosfet model and its applications
  to cmos inverter delay and other formulas,'' \emph{IEEE JSSC}, vol.~25,
  no.~2, pp. 584--594, Apr 1990.

\bibitem{ITRS}
\BIBentryALTinterwordspacing
S.~Association. International technology roadmap for semiconductors 2005.
  [Online]. Available: \url{http://public.itrs.net/Links/2005ITRS/Home2005.htm}
\BIBentrySTDinterwordspacing

\bibitem{audibert2010}
\BIBentryALTinterwordspacing
J.-Y. Audibert and S.~Bubeck, ``{Best Arm Identification in Multi-Armed
  Bandits},'' in \emph{{COLT - 23th Conference on Learning Theory - 2010}},
  Haifa, Israel, Jun. 2010, p. 13 p. [Online]. Available:
  \url{https://hal-enpc.archives-ouvertes.fr/hal-00654404}
\BIBentrySTDinterwordspacing

\bibitem{Gabillon2012}
\BIBentryALTinterwordspacing
V.~Gabillon, M.~Ghavamzadeh, and A.~Lazaric, ``Best arm identification: A
  unified approach to fixed budget and fixed confidence,'' in \emph{Advances in
  Neural Information Processing Systems}, F.~Pereira, C.~Burges, L.~Bottou, and
  K.~Weinberger, Eds., vol.~25.\hskip 1em plus 0.5em minus 0.4em\relax Curran
  Associates, Inc., 2012. [Online]. Available:
  \url{https://proceedings.neurips.cc/paper/2012/file/8b0d268963dd0cfb808aac48a549829f-Paper.pdf}
\BIBentrySTDinterwordspacing

\bibitem{gabillon2011multi}
V.~Gabillon, M.~Ghavamzadeh, A.~Lazaric, and S.~Bubeck, ``Multi-bandit best arm
  identification,'' \emph{Advances in Neural Information Processing Systems},
  vol.~24, 2011.

\end{thebibliography}

\vspace{12pt}
\end{document}